# Interpretable machine-learning for predicting molecular weight of PLA based on artificial bee colony optimization algorithm and adaptive neuro-fuzzy inference system

Amir Pouya Masoumi
*Centre for Mathematical Modelling and Intelligent Systems for Health and Environment (MISHE), Atlantic Technological University*
Sligo, Ireland
Amirpouya.Masoumi@atu.ie

Leo Creedon
*Centre for Mathematical Modelling and Intelligent Systems for Health and Environment (MISHE), Atlantic Technological University*
Sligo, Ireland
Leo.Creedon@atu.ie

Ramen Ghosh
*Centre for Mathematical Modelling and Intelligent Systems for Health and Environment (MISHE), Atlantic Technological University*
Sligo, Ireland
Ramen.Ghosh@atu.ie

Nimra Munir
*Centre for Mathematical Modelling and Intelligent Systems for Health and Environment (MISHE), Atlantic Technological University*
Sligo, Ireland
nimra.munir@research.atu.ie

Ross McMorrow
*Centre for Mathematical Modelling and Intelligent Systems for Health and Environment (MISHE), Atlantic Technological University*
Sligo, Ireland
ross.mcmorrow@atu.ie

Marion McAfee
*Centre for Mathematical Modelling and Intelligent Systems for Health and Environment (MISHE), Atlantic Technological University*
Sligo, Ireland
Marion.McAfee@atu.ie

**Abstract**—This article discusses the integration of the Artificial Bee Colony (ABC) algorithm with two supervised learning methods, namely Artificial Neural Networks (ANNs) and Adaptive Network-based Fuzzy Inference System (ANFIS), for feature selection from Near-Infrared (NIR) spectra for predicting the molecular weight of medical-grade Polylactic Acid (PLA). During extrusion processing of PLA, in-line NIR spectra were captured along with extrusion process and machine setting data. With a dataset comprising 63 observations and 512 input features, appropriate machine learning tools are essential for interpreting data and selecting features to improve prediction accuracy. Initially, the ABC optimization algorithm is coupled with ANN/ANFIS to forecast PLA molecular weight. The objective functions of the ABC algorithm are to minimize the root mean square error (RMSE) between experimental and predicted PLA molecular weights while also minimizing the number of input features. Results indicate that employing ABC-ANFIS yields the lowest RMSE of 282 Da and identifies four significant parameters (NIR wavenumbers 6158 cm$^{-1}$, 6310 cm$^{-1}$, 6349 cm$^{-1}$, and melt temperature) for prediction. These findings demonstrate the effectiveness of using the ABC algorithm with ANFIS for selecting a minimal set of features to predict PLA molecular weight with high accuracy during processing.

I. INTRODUCTION

Spectroscopic instruments such as NIR, Raman, and UV-Vis, which can detect changes in molecular bond activity have proven to be valuable assets for real-time monitoring of quality parameters in material processing, such as degradation, mechanical properties, and particle size. Raman and NIR spectroscopy have been shown to exhibit sensitivity to degradation of molecular weight during polymer processing [1]. However, spectra can be difficult to interpret manually and can also present challenges for machine learning algorithms due to having large numbers of features, many of which are typically highly correlated. Having unnecessary and repeated factors can negatively impact predictive models by increasing the calculation time and complexity and decreasing the accuracy. Therefore, obtaining a smaller subset of the most relevant features is vitally important. This practice saves time in computation, while preventing overfitting issues in prediction models. High dimensionality has been a significant challenge in recent years for data mining, pattern recognition, computer vision, and machine learning, in many fields. Feature selection and feature extraction are central strategies to overcome this challenge by aiming to eliminate redundant features and enhance learning algorithm performance. While latent variable methods of dimension reduction such as Principal Component Analysis (PCA) are widely used, these are not optimal for nonlinear relationships and further obscure the interpretability compared to selection of a subset of original features. Feature selection methods have been introduced to eliminate irrelevant and redundant features to achieve the smallest, most powerful subset of features in order to reduce the computation time and improve the prediction accuracy simultaneously. Feature selection techniques have several branches including filter methods [2] [3], wrapper methods [3], embedded methods, subspace learning, sparse representation learning, information theory, evolutionary computational algorithms, and reinforcement learning [4]. Liu *et al.* [5] have presented the Relief F-based feature selection algorithm for cancer histology (REDFS), aiming to improve prediction accuracy in three types of cancer including gastric, thyroid, and breast cancers. The algorithm involves two stages, namely feature selection and training prediction. It integrates difference coefficients to enhance stability. Outcomes from experiments on three cancer datasets show superior classification performance, with the algorithm successfully identifying breast cancer subtypes through clustering analysis. Survival analysis confirms the significance of the classified subtypes and signature genes demonstrate the ability to predict cancer development over time. Hasanzadeh *et al.* [6] proposed a hybrid method for Parkinson's disease diagnosis, focusing on gender-specific speech traits and utilizing Statistical-Based Feature Score and Classification-Based Feature Score. The method uses fewer features than previous studies and achieved 86% and 84% accuracy for men and women respectively. Zhang *et al.* [7] presented a novel method named Distance Correlation Gain-Network (DCG-Net) for biological data analysis. DCG-Net measures interactive effects between pairwise features, constructing a molecular interactive effect network in physiological and pathological changes. Comparative evaluations on ten datasets demonstrate DCG-Net's superior performance in classification accuracy, sensitivity, and specificity across different classifiers including t-test, ReliefF, support vector machine-recursive feature elimination regressions. Evolutionary computational and machine learning methods are recognized as branches of feature selection and feature extraction. Osco *et al.* [8] investigated the impact of pre-processing techniques on the performance of machine learning algorithms in classifying vegetation insect damage using hyperspectral data. Various types of pre-processing algorithms, including baseline removal, smoothing, derivatives, standard normal variate (SNV), Extra-Tree algorithm, multiplicative scatter correction (MSC), and principal component analysis (PCA), were examined. The results indicate that the Extra-Tree algorithm achieved the highest accuracy of 93.68%. Dyar *et al.* [9] employed machine learning algorithms instead of relying solely on spectral features for the classification of asteroid belts. Comparison results demonstrated that logistic regression exhibited the highest accuracy and could distinguish eight robust groups of meteorite classes. Hickey *et al.* [10] proposed the Locus Algorithm to enhance differential photometry by optimizing the selection of reference stars. Supervised machine learning models, including Support Vector Machine Regression (SVR), Random Forest (RF), Gradient Boosting Machine (GBM), and regularized linear models, were trained and evaluated. The results indicated that SVR demonstrated the best performance. The SVR model achieved an $R^2$ coefficient of 0.62, suggesting improved performance for the Locus Algorithm. Mulrennan *et al.*[11] used in-process NIR spectroscopy, pressure, and temperature data as inputs for SVR and RF regressions with and without dimension reduction via PCA to estimate the mechanical strength of an extruded PLA product. Their findings reveal that PCA was an essential dimension reduction step for good accuracy and that employing RF yielded a superior root mean square error compared to SVR. A disadvantage of the approach is the black box nature of the predictions.

Molecular weight is considered a critical parameter in PLA processing that impacts mechanical properties, processability, thermal stability, degradation rate, and other material properties. Understanding and monitoring molecular weight can lead to the development of polymer formulations with desired characteristics for specific applications in various industries. Additionally, one of the challenging aspects of estimating molecular weight in PLA is the high dimensionality of the inputs. In a previous investigation [1], the recursive feature elimination (RFE) method was employed for feature selection and predicting the molecular weight of PLA. Using Monte Carlo cross-validation, a mean $R^2$ of 0.86% was achieved for an RFE-RF molecular weight predictive model. The objective of this study is to explore alternative machine learning algorithms, such as Artificial Bee Colony and supervised learning algorithms to further refine feature selection, enhance interpretability, and improve the accuracy of molecular weight prediction. In this paper, firstly, the importance of processing medical-grade PLA, along with the measurement setup and dataset, is introduced. Subsequently, two supervised learning techniques, namely ANN and ANFIS, along with the Artificial Bee Colony algorithm are described. Finally, the significant features utilized for predicting the molecular weight of PLA are identified with a discussion of performance.

## II. Overview of PLA and Experimental Methodology

Polylactic acid (PLA) is a biodegradable and thermoplastic polyester derived from renewable resources such as corn starch or sugarcane. It is considered one of the environmentally friendly alternatives to traditional petroleum-based plastics because it is compostable and biodegradable under the right conditions. PLA is widely used in various applications, including food packaging, textiles, medical devices, and 3D printing. One of the most important properties of PLA is the molecular weight, which influences mechanical properties, process characteristics, degradation rate, and thermal properties. The molecular weight can be degraded by processing, and hence monitoring and control of the process to ensure the molecular weight is sufficient is vitally important for critical applications such as medical devices.

Medical-grade PLA undergoes purification procedures, eliminating impurities such as catalyst traces to meet FDA regulations for medical applications. The study utilized a dataset related to the extrusion of medical grade PLA, comprising 63 experiments conducted in the development of a commercial process for fabrication of tissue scaffold medical device. Each experiment varied temperature settings, screw speed, and feed rate, while in-process sensors captured data on the melt temperature, pressure and NIR spectra of the melt, resulting in a total of 512 input features and 63 samples. Molecular weight analysis was conducted using high-performance liquid chromatography on samples prepared in chloroform [1]. NIR data were collected for wavenumbers 4000–7500 cm$^{-1}$ at a resolution of 4 cm$^{-1}$. From the NIR data, only wavenumber ranges from 6101 to 6599 cm$^{-1}$ (499 wavenumbers -see Figure 1) were included as input features in the model, due to poor signal-to-noise ratio in other regions. More information regarding the experiments can be seen in [1] and the dataset is freely available for download [12].

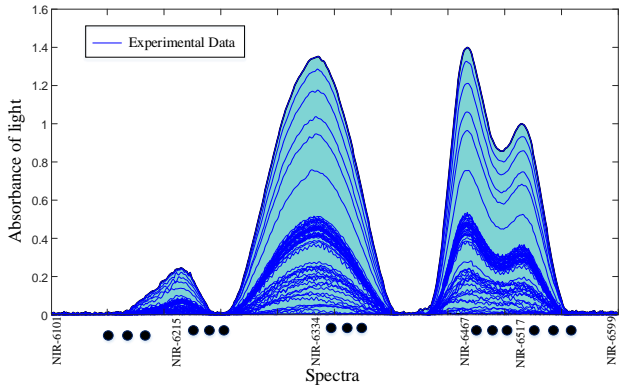

Fig. 1 Absorbance of light with NIR spectra (6101-6599 cm$^{-1}$)

## III. OVERVIEW OF OPTIMIZATION AND SUPERVISED LEARNING METHODS

Popular supervised learning algorithms include linear regression, logistic regression, support vector machines, decision trees, random forests, artificial neural networks, and adaptive network-based fuzzy inference system. These algorithms are widely used for tasks such as classification, regression [13], and prediction [14].

Population-based metaheuristic optimization algorithms draw inspiration from various natural phenomena such as chromosomes, fish, bees, ants, birds, and fireflies. Their primary function is to search for optimal unknown parameters within a system, aiming to either minimize or maximize an objective function. A population of solutions is initialized and evaluated for fitness (performance with respect to the objective function), and different strategies are used to iterate towards identification of an optimal solution. Several metaheuristic optimization algorithms have garnered attention in artificial intelligence research including Genetic Algorithms, Particle Swarm Optimization, Artificial Bee Colony (ABC), and Ant Colony. During initialization, the initial population is randomly or uniformly generated. The main loop involves iteratively seeking the optimal parameter value based on the algorithm's fitness function, this is terminated on meeting some pre-defined criteria.

### A. Artificial Bee Colony

The Artificial Bee Colony algorithm belongs to a group of swarm intelligence methods that mimic the foraging behaviours of honey bee swarms. It has found successful applications in various domains such as optimization, signal processing, and controller design due to its efficient search capabilities. In this algorithm, the search process is divided into three stages: employed bee, onlooker, and scout stages. Initially, all artificial bees are dispatched randomly to explore potential solutions (scout), after which they are categorized into employed and onlooker bees based on the amount of 'nectar'. Here, nectar represents the fitness of a solution, and the ABC optimization algorithm tries to maximize the overall nectar content by iteratively selecting and improving solutions until a performance or convergence threshold is reached.

Employed bees with higher nectar levels proceed to search for solutions in their neighbourhood and share their findings, while onlookers probabilistically select promising sources to explore based on the information from the employed bees. If a solution fails to improve over several iterations, it is discarded, and the corresponding employed bee transitions into a scout to explore new options randomly. This iterative process continues until the best solution is found or termination conditions are met. The algorithm effectively balances exploration and exploitation, drawing parallels to the division of labour and communication strategies among honey bee classes in nature [15]. The standard flowchart of ABC optimization algorithm can be seen in Figure 2.

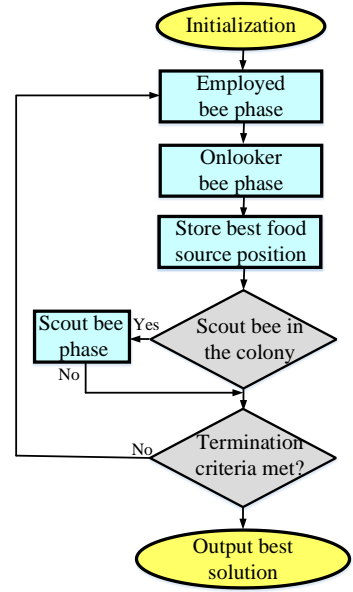

Fig. 2 Flow chart of Artificial Bee Colony optimization algorithm

### B. Artificial Neural Network

The ANN algorithm typically comprises three distinct layers. Each layer consists of multiple neurons. The first layer, known as the input layer, contains a predetermined number of neurons corresponding to the number of model inputs. The second layer, termed the hidden layer, assumes an essential role in processing inputs by adjusting weights and biases to generate outputs. The number of neurons in the hidden layer is manually determined through trial and error, considering the data complexity. Lastly, the third layer, the output layer, consists of neurons equal in number to the desired output quantity (see Figure 3).

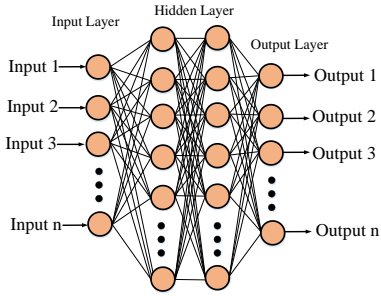

Fig.3 Artificial Neural Network structure

### C. Adaptive Network-based Fuzzy Inference System

ANFIS, also known as the Adaptive Network-based Fuzzy Inference System, operates within the framework of adaptive networks which encompass various types of feedforward neural networks capable of supervised learning. ANFIS serves as the basis for constructing a collection of fuzzy if-then rules complete with appropriate membership functions to generate specified input-output pairs. The ANFIS architecture consists of five layers, each serving a specific function in the inference process. These layers are the fuzzification layer, rule layer, normalization layer, defuzzification layer, and output layer. The input variables are first fuzzified using linguistic terms and membership functions in the fuzzification layer. Then fuzzy if-then rules are applied in the rule layer to represent the system's behaviour. The normalized firing strengths of the rules are computed in the normalization layer, followed by defuzzification to obtain the crisp output in the defuzzification layer [13]. Finally, the overall output is computed in the output layer. A flowchart of the ANFIS can be seen in Figure 4.

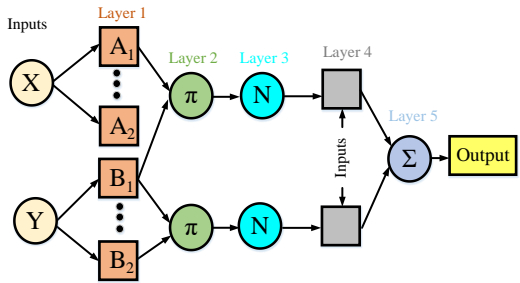

Fig. 4 Adaptive Network-based Fuzzy Inference System structure.

## IV. FEATURE SELECTION AND PREDICTIVE MODELLING OF PLA MOLECULAR WEIGHT

Feature selection is a critical strategy aimed at reducing system complexity, enhancing model interpretability, and minimizing computational costs. In this study, the dataset comprises 512 features. Analysing all these features is challenging, thus the use of feature selection methods becomes essential to identify the most relevant inputs for predicting molecular weight. Various approaches, including the filter method, wrapper method, and machine learning techniques, are available for feature selection. Here, the integration of an optimization algorithm with ANN and ANFIS is considered as a feature extraction and regression method for predicting the target. This section outlines the utilization of the Artificial Bee Colony (ABC) algorithm to identify the essential features for predicting molecular weights. Integrating ABC into the ANN/ANFIS regressions aims to determine the minimal number of features needed to minimize the mean square error between experimental and predicted molecular weights. ABC systematically identifies and eliminates features that do not significantly impact the prediction, resulting in a set of features comprising 499 spectra, 4 machine set-up features relating to extrusion die and fibre spooling, temperature in each zone, melt temperature in the extruder die, and feed rate. The objective function of the ABC optimization algorithm is to minimize the root mean square error (RMSE) between experimental data and predicted results for the molecular weight of PLA, and also minimize the number of input features. In fact, ABC aims to minimize two cost functions concurrently. The ABC algorithm is executed with a population size of 50 over 25 iterations. With respect to the ANN and ANFIS, eight structures are considered as can be seen in Table I. Tangent hyperbolic function is used for ANN and Gaussian membership function is used for ANFIS. Figure 5 illustrates the block diagram depicting the utilization of the ABC optimization algorithm alongside ANN/ANFIS supervised learning algorithms for feature selection.

TABLE I. DIFFERENT STRUCTURES FOR REGRESSION

| Case | Regressions | Parameters in Hidden layer | Output |
|------|-------------|---------------------------|--------|
| 1 | ANFIS | 7 membership function | constant |
| 2 | ANFIS | 7 membership function | linear |
| 3 | ANFIS | 13 membership function | constant |
| 4 | ANFIS | 13 membership function | linear |
| 5 | ANN | 10 neurons and one hidden layer | linear |
| 6 | ANN | 20 neurons and one hidden layer | linear |
| 7 | ANN | 30 neurons and one hidden layer | linear |
| 8 | ANN | 40 neurons and one hidden layer | linear |

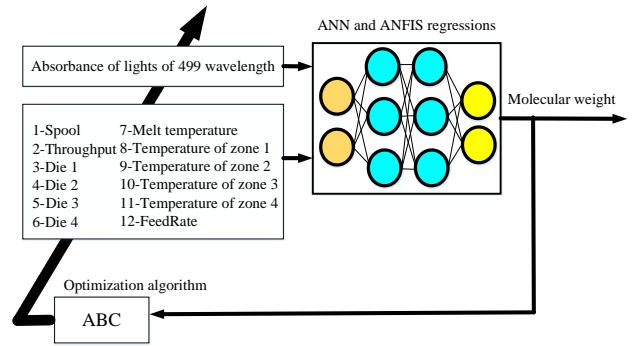

Fig. 5 Block diagram of ABC applied to ANN/ANFIS regression models for predicting molecular weight of PLA

The sample size is imbalanced relative to the number of features. To ensure the effectiveness of regression results, the K-fold cross-validation method was employed. The data is divided into 5 segments, and the root mean square error (RMSE) and coefficient of determination ($R^2$) are computed for each fold. Subsequently, the mean RMSE and mean $R^2$ values across all segments are considered as benchmarks for comparing the final accuracy with various results. For each model, the standard deviation of the errors over all folds is also calculated as a measure of the model robustness. Figures 6 and 7 illustrate the average RMSE for different numbers of features with varying structures of ANFIS and ANN. It is evident that the average RMSE for ANFIS is lower than for ANN for all structures. Table II presents the top four models for predicting the molecular weight of PLA, while Figure 8 displays the coefficient of determination of each fold for the best-case model. ANFIS with 13 membership functions and linear output yields the lowest RMSE with selected input features: NIR wavenumbers 6158 cm$^{-1}$, 6310 cm$^{-1}$, 6349 cm$^{-1}$, and the melt temperature. This model has a mean RMSE of 281.83 Da and an average coefficient of determination of 0.96.

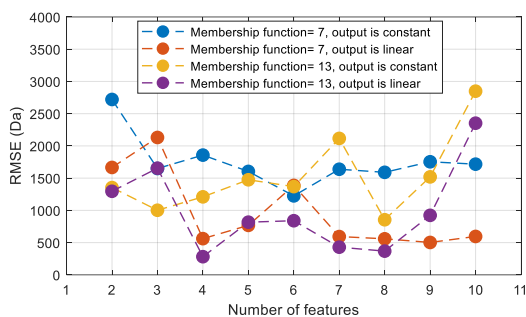

Fig. 6 RMSE between experimental data and predicted results using ABC-ANFIS

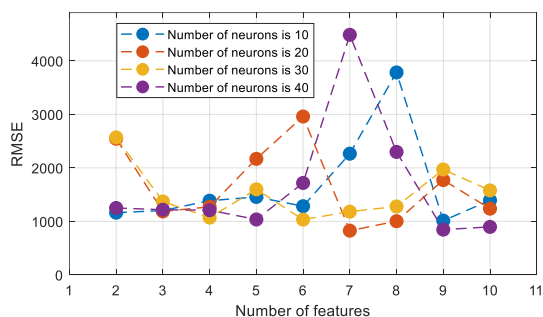

Fig. 7 RMSE between experimental data and predicted results using ABC-ANN

TABLE II. BEST RESULTS BASED ON ABC-ANN/ANFIS

| Supervised learning algorithm | Structure of regression model | Mean RMSE (Da) | Selected features | Standard deviation of the error (Da) |
|---|---|---|---|---|
| ANFIS | MF=13 Output linear | 281.83 | NIR 6158 cm$^{-1}$, NIR 6310 cm$^{-1}$, NIR 6349 cm$^{-1}$, Melt temperature | 345.83 |
| ANFIS | MF=13 Output linear | 368.22 | NIR 6203 cm$^{-1}$, NIR 6494 cm$^{-1}$, NIR 6291 cm$^{-1}$, NIR 6392 cm$^{-1}$, | 412.40 |
| | | | NIR 6456 cm$^{-1}$, NIR 6218 cm$^{-1}$, NIR 6401 cm$^{-1}$, Melt temperature | |
| ANFIS | MF=13 Output linear | 429.48 | NIR 6419 cm$^{-1}$, NIR 6300 cm$^{-1}$, NIR 6263 cm$^{-1}$, NIR 6480 cm$^{-1}$, NIR 6497 cm$^{-1}$, NIR 6301 cm$^{-1}$, Melt temperature | 472.42 |
| ANFIS | MF=7 Output linear | 502.64 | NIR 6406 cm$^{-1}$, NIR 6543 cm$^{-1}$, NIR 6165 cm$^{-1}$, NIR 6443 cm$^{-1}$, NIR 6427 cm$^{-1}$, NIR 6148 cm$^{-1}$, NIR 6395 cm$^{-1}$, NIR 6503 cm$^{-1}$, Melt temperature | 708.72 |

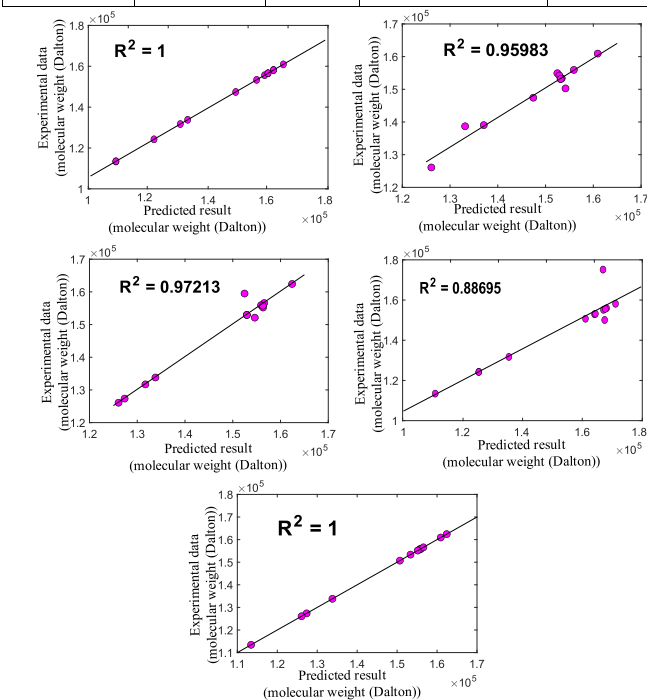

Fig. 8 Coefficient of determination between experimental data and predicted results for ANFIS with 13 membership functions and linear output for 5-fold cross validation

CONCLUSION

In this paper, the Artificial Bee Colony optimization algorithm is applied in combination with supervised learning algorithms (ANN and ANFIS) to predict the molecular weight of medical-grade PLA. Predicting the molecular weight presents a challenging task due to the dataset containing 512 input features, making interpretation of all features difficult. Consequently, a feature selection approach is employed to accurately predict the molecular weight by eliminating redundant features. In this study, the ABC optimization algorithm tries to simultaneously minimize two cost functions: 1) minimizing the RMSE between experimental data and simulation results in molecular weight, and 2) minimizing the number of features or inputs. According to the results, ANFIS yields better accuracy and reliability compared to ANN. The best RMSE achieved is 282Da using ANFIS with 13 membership functions and linear output and with the selection of only four features from all sensor and process data. The significant inputs selected by ABC are NIR wavenumbers 6158 $cm^{-1}$, 6310 $cm^{-1}$, 6349 $cm^{-1}$, and melt temperature. In contrast, the previous study using an RFE-RF algorithm selected 9 features to predict molecular weight. Additionally, the $R^2$ of the ANFIS model in this study is approximately 0.96 in 5-fold cross-validation, compared to an $R^2$ of around 0.86 in the previous study. However, it should be noted that the earlier work utilized the Monte Carlo cross-validation method with 100 random train-test splits so this is not a direct comparison. The feature selection approach helps in reducing the high-dimensional dataset to a smaller set of inputs, thereby enhancing interpretability. The ABC-ANFIS approach shows promise for developing accurate and simple models from high-dimensional data and outperforms recent work using Recursive Feature Elimination. However, model robustness and computation time are also relevant performance measures in predictive modelling and future work will be dedicated to investigating ABC with other types of supervised learning algorithms to deal with similarly complex datasets.

ACKNOWLEDGMENT

The research leading to these results has received funding from Science Foundation Ireland under grant number 21/FFP-A/9152. For the purpose of Open Access, the authors have applied a CC BY public copyright license to any Author Accepted Manuscript version arising from this submission.